\documentclass[twocolumn,superscriptaddress,showpacs]{revtex4-1}

\usepackage{setspace}
\usepackage{bm}
\usepackage{bbm}
\usepackage[T1]{fontenc}
\usepackage{amsmath}
\usepackage{latexsym}
\usepackage{amssymb}
\usepackage{float}
\usepackage{graphicx}           
\usepackage{color}
\usepackage{amsmath}
\usepackage{amssymb}
\usepackage{amsthm}
\usepackage{url}
\usepackage{braket}
\usepackage{hyperref}
\usepackage{bbold,bbm}
\usepackage[mathscr]{euscript}
\usepackage{mathtools}
\usepackage{verbatim}
\usepackage{bbold,bbm}
\usepackage{times}
\usepackage[T1]{fontenc}
\hypersetup{
    colorlinks=true,
    linkcolor=blue,
    filecolor=magenta,      
    urlcolor=cyan,
}
\newcommand{\nbox}[2][9]{\hspace{#1pt} \mbox{#2} \hspace{#1pt}}

\newcommand{\RNum}[1]{\uppercase\expandafter{\romannumeral #1\relax}}
\usepackage{soul}

\newcommand{\sH}{\mathscr{H}}

\newcommand{\beq}[0]{\begin{equation}}
\newcommand{\eeq}[0]{\end{equation}}

\newcommand{\one}{\leavevmode\hbox{\small1\normalsize\kern-.33em1}}

\def\be{\begin{equation}}
\def\ee{\end{equation}}
\def\ben{\begin{eqnarray}}
\def\een{\end{eqnarray}}
\def\eea{\end{array}}
\def\bea{\begin{array}}

\newcommand{\Tr}[1]{\mathrm{Tr}#1}
\newcommand{\bei}{\begin{itemize}}
\newcommand{\eei}{\end{itemize}}

\newcommand{\I}{\mathbbm{1}}

\renewcommand{\emph}[1]{\textbf{#1}}

\makeatletter
\newtheorem*{rep@theorem}{\rep@title}
\newcommand{\newreptheorem}[2]{%
\newenvironment{rep#1}[1]{%
 \def\rep@title{#2 \ref{##1}}%
 \begin{rep@theorem}}%
 {\end{rep@theorem}}}
\makeatother

\theoremstyle{plain}
\newtheorem*{thm*}{Theorem}
\newreptheorem{thm}{Theorem}

\theoremstyle{definition}

\theoremstyle{remark}

\usepackage[T1]{fontenc}
\begin{document}

\title{Device-independent certification of maximal randomness from pure entangled two-qutrit states using non-projective measurements}

\author{Jakub Jan Borkała}
\author{Chellasamy Jebarathinam}
\author{Shubhayan Sarkar}
\author{Remigiusz Augusiak}
\affiliation{Center for Theoretical Physics, Polish Academy of Sciences, Aleja Lotnik\'{o}w 32/46, 02-668 Warsaw, Poland}

\begin{abstract}

While it has recently been demonstrated how to certify the maximal amount of randomness from any pure two-qubit entangled state in a device-independent way [E. Woodhead \textit{et al.}, \href{https://journals.aps.org/prresearch/abstract/10.1103/PhysRevResearch.2.042028}{Phys. Rev. Research \textbf{2}, 042028(R) (2020)}], the problem of optimal randomness certification from entangled states of higher local dimension remains open. Here we introduce a method for device-independent certification of the maximal possible amount of $2\log_23$ random bits using pure bipartite entangled two-qutrit states and extremal nine-outcome general non-projective measurements. To this aim, we exploit the extended Bell scenario introduced recently in [S. Sarkar \textit{et al.}, \href{https://arxiv.org/abs/2110.15176}{arXiv:2110.15176}], which combines a device-independent method for certification of the full Weyl-Heisenberg basis 
in three-dimensional Hilbert spaces together with a one-sided device-independent method for certification of two-qutrit partially entangled states. 

\end{abstract}

\maketitle


\section{Introduction}
The intrinsic randomness of quantum theory manifested in the outcomes of quantum measurement is one of the most intriguing features of quantum mechanics \cite{Bera_2017}. Even more remarkable is the fact that quantum technologies allow us to generate certifiable randomness with an unprecedented level of security \cite{AM16}. Protocols designed for randomness certification ensure both the generation of completely random bits and their privacy, which for instance, introduces new possibilities in designing protocols for tasks like quantum cryptography and quantum key distribution \cite{Schwonnek2021}.   

Since the pioneering works on randomness certification \cite{Pironio2010} (see also Ref. \cite{AM16}), significant progress has been made both in theoretical and experimental aspects \cite{Bierhorst2018,Liu2018,MYC+16,HGJ17}. 
It was shown, for instance, in Ref. \cite{SSK+19} that maximal violation of the Salavrakos-Augusiak-Tura-Wittek-Acín-Pironio (SATWAP) Bell inequality \cite{SAT+17} enables self-testing the maximally entangled state of two-qudits of arbitrary local dimension, which in turn allows certifying $\log_2d$ bits of randomness by using projective measurements. On the other hand, we know that non-projective measurements, also known as positive-operator valued measures (POVM), can be used to generate more randomness in a given dimension than projective ones. The intuitive reason behind this is the existence of extremal $d^2$-outcome non-projective measurements in $d$-dimensional Hilbert spaces, which consequently might give rise to $2\log_2d$ random bits \cite{APP05}. In fact, a method for certification of two bits of local randomness in dimension two by exploiting such non-projective measurements was introduced in Ref. \cite{APV+16}. 

Similar research was conducted in Ref. \cite{ABD+18}, where the authors exploited  Gisin's elegant Bell inequality \cite{Gis09} instead of the Clauser-Horne-Shimony-Holt (CHSH) inequalities used in Ref. \cite{APV+16}. Later, in Ref. \cite{WKB+20} it was shown how to certify the maximal amount of local randomness independently of the degree of entanglement of two-qubit states.

While significant progress has been made in understanding the possibility of device-independent (DI) randomness certification from entangled states of the lowest possible dimension, higher dimensional scenarios remain mostly unexplored; see nevertheless the recent work \cite{TFR+21} presenting an approach in which by using symmetric informationally complete POVMs one can obtain more than $\log_2d$ bits of local randomness from the maximally entangled two-qudit state of the local dimension up to $d=7$. 




However, it remains an open and highly nontrivial problem whether it is possible to device-independently certify the maximal amount of $2\log_2d$ bits of randomness 
by performing measurements on quantum systems of dimension $d$ for any finite $d$. Another interesting direction to explore is whether the maximal amount of randomness can be certified independently of the degree of entanglement of states used in the protocol. We provide a positive answer to the first problem and a partial solution to the second one in dimension three; that is, we show how to certify  $2\log_23$ bits of randomness by performing local non-projective measurements on a well-defined subset of pure bipartite entangled states in a fully device-independent way. 

In our work, we use the family of Bell inequalities proposed in Ref. \cite{KST+19} that allows for self-testing the two-qutrit maximally entangled state as well as three mutually unbiased bases (MUBs) per party. We extend the self-testing proof of \cite{KST+19} to certify, up to the transposition equivalence, all the Weyl-Heisenberg (W-H) operators acting on three-dimensional Hilbert spaces. Our approach for W-H basis certification is inspired by Ref. \cite{WKB+20} and is based on simultaneous maximal violation of the Bell inequality from Ref. \cite{KST+19} by two appropriately selected sets of measurements. This way, we can self-test a complete set of four MUBs, which allows us to construct eight W-H operators in dimension three. As a consequence of certifying the complete W-H basis, we can characterise any measurement acting on three-dimensional Hilbert space in terms of basis elements \cite{BBR+02}. 
Let us note that our self-testing statements for measurements are always up to 
the standard equivalences, but also up to the tranposition equivalence.

The structure of our article is the following. In the preliminaries, we first present the scenario used in our work. Next, we review the Bell inequality from \cite{KST+19}, with slight modifications that are necessary for presenting our results. Afterwards, we also review the steering inequality introduced in Ref. \cite{SBJ+21} which, together with the Bell inequality \cite{KST+19} enables certification of any pure bipartite entangled state of local dimension three. In the second part, where we present our results, we provide a method for DI certification of the full W-H basis. Then, we present the main result of our work, which is proof for DI certification of a maximal amount of local randomness from  pure entangled states in dimension three. Finally, we recall the construction of extremal qutrit POVM of Ref. \cite{SBJ+21} for a significant subset of partially entangled states that can be used for optimal randomness certification.

\section{Preliminaries}

\subsection{Scenario}

Since we are concerned with the device-independent certification of randomness, we consider an adversarial Bell scenario, consisting of two parties, Alice and Bob, and an adversary, Eve. Alice and Bob cannot trust their devices in this scheme because Eve, a malicious eavesdropper, could have full control of all of their resources. An example of Eve's strategy might be to use extra dimensions of the Hilbert space hidden in the devices to learn about the results of Alice's and Bob's measurements. 
Eve may also try to entangle with the subsystems of our protagonists Alice and Bob and create correlations, which will allow her to obtain some information on the outputs of the experiment. Nevertheless, the strength of the randomness certification techniques lies in the possibility to prove that, despite any attacks, Eve cannot learn anything about the results of Alice's and Bob's measurements. Security of the protocol is demonstrated if they both observe strong correlations in their measurement statistics, i.e. correlations which exhibit the maximal quantum violation of a given Bell inequality. 

We construct the following scenario to certify randomness from pure bipartite entangled states of local dimension three in a device-independent way. Alice and Bob perform local measurements on their quantum subsystems, labelled by $A$ and $B$, which they receive from the preparation device $\mathcal{P}$ operated by Charlie and consisting of two inputs $p=1,2$. Preparation $\mathcal{P}_1$ corresponds to preparing  a state $\rho^1_{AB}$ and $\mathcal{P}_2$ a state $\rho^2_{AB}$. Both preparations can be purified as $\ket{\psi^1}_{ABE}$ and $\ket{\psi^2}_{ABE}$ respectively. Charlie can freely choose the input of the preparation device.

\begin{center}
     \begin{figure}[h!]
    \includegraphics[scale=0.23]{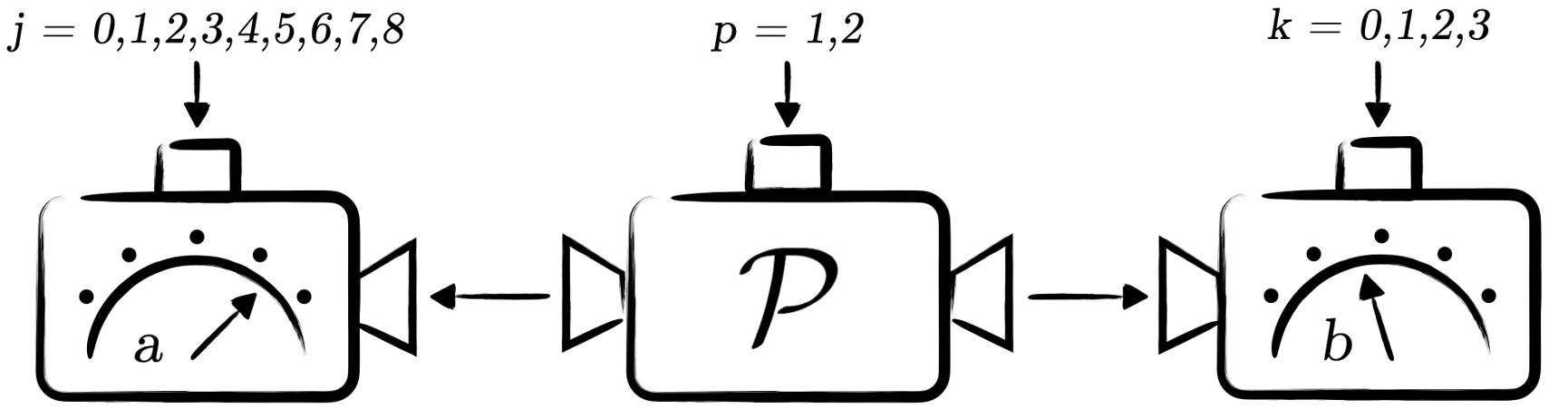}
    \caption{\textbf{Randomness certification scenario for $d=3$.} 
    Alice and Bob have access to untrusted devices, and they apply measurements $F_j$ and $G_k$, respectively. The preparation box distributes two different bipartite states $\rho^1_{AB}$ for the preparation $\mathcal{P}_1$ and $\rho^2_{AB}$ for the preparation $\mathcal{P}_2$. After collecting the measurements statistics $\vec{p}$ one can device-independently certify Bob's measurements based on Alice's inputs $j=0,1,2,3,4,5$ and Bob's inputs $k=0,1,2,3$. The preparation $\mathcal{P}_2$ can be certified to be any pure entangled state of local dimension $3$ with inputs $j=6,7$ and $k=0,1$.
    Randomness certification is based on the preparation $\mathcal{P}_2$ and $9$-outcome measurement corresponding to the input $j=8$.}
    \label{fig}
\end{figure}
\end{center}

Alice's device has nine inputs labelled by $j=0,\dots,8$ and Bob's device has four inputs labelled by $k=0,1,2,3$. The first eight measurements of Alice and all measurements of Bob results in three outputs, labelled by $a$ for Alice and $b$ for Bob such that $a,b=0,1,2$. The ninth measurement on Alice's side corresponding to $j=8$ is a nine-outcome measurement. We employ this additional measurement to certify randomness from its outcomes. A schematic representation of our scenario is presented in Fig. \ref{fig}. It is necessary here to assume that the measurements are independent of the input of the preparation device.

Alice, Bob and Charlie now collect statistics for each input and the corresponding outputs, which allows them to reconstruct the probability distribution  $\vec{p}=\{p(a,b|j,k,p)\}$, where $p(a,b|j,k,p)$ is the probability that outcomes $a$ and $b$ are obtained when performing measurements $j$ and $k$ on the subsystems $A$ and $B$ given the prepartion $p$. Using this scenario, one can first certify the full Weyl-Heisenberg basis (section \ref{sec:H-W basis}), then any entangled state of local dimension $d=3$, and, finally, the optimal amount of randomness from entangled states of local dimension $d=3$ (section \ref{sec:randomness}).

Let us now discuss potential Eve's strategies that need to be taken into account to ensure that the generated randomness is not accessible by her. For example, suppose Alice wants to generate randomness from the outcomes of one of her measurements. Then the aim for Eve is to guess Alice's outputs with the highest possible probability. To do so, Eve can prepare Alice's and Bob's systems in any way compatible with the given statistics $\vec{p}$ by using quantum resources while remaining undetected. Therefore we can characterize Eve's strategy $S$ applied for the attack using four major points:
\begin{enumerate}
   \item  Eve, for instance, may know the input of the preparation device $p$ and the inputs of Alice and Bob $x,y$, but she cannot change them.
    \item  
   Eve might possess some subsystem $E$ correlated with both the parties. Consequently, the state shared among Alice and Bob is defined by $\rho_{AB}=\Tr_E(\rho_{ABE})$, where $\rho_{ABE}\in \sH_A\otimes \sH_B\otimes\sH_E $ denotes the state shared among Alice, Bob and Eve such that the local Hilbert spaces can be of any arbitrary dimension.
    \item Eve might have control over Alice's and Bob's measurement devices, that is  POVM $F_j=\{F_{a|j}\}$ on $\sH_A$ and POVM $G_k=\{G_{b|k}\}$ on $\mathcal{H}_B$ respectively.
    \item Eve's device is characterised by POVM $Z=\{Z_a\}$ on $\mathcal{H}_E$. The probability of obtaining outcome $a$ from measurement performed by Eve on her share of the joint state $\rho_{ABE}$ is the best guess of Alice's outcome $a$.
\end{enumerate}
Since there is no restriction on Eve's subsystem, we can safely assume here that $\rho_{ABE}$ is pure and write $\rho_{ABE}=\ket{\Psi_{ABE}}\!\!\bra{\Psi_{ABE}}$. 
Eve's influence remains undetected if it cannot be observed in the statistics $\vec{p}$ obtained by Alice and Bob, i.e. 
\begin{equation}
\label{qcorr}
p(ab|jk)=\bra{\Psi_{ABE}} F_{a|j} \otimes G_{b|k} \otimes \mathbb{1}_E \ket{\Psi_{ABE}}.
\end{equation}
%
Now let us define the local guessing probability, that is the probability that Eve's guess agrees with Alice's output 
\begin{equation}\label{LGpr}
    G(j,\vec{p})=\sup_{S\in S_{\vec{p}}}\sum_{a}\bra{\Psi_{ABE}}\left( F_{a|j} \otimes \I_B \otimes Z_a \right)\ket{\Psi_{ABE}},
\end{equation}
where the supremum is taken over all strategies $S_{\vec{p}}$, consisting of the shared state $\rho_{ABE}$, Bob's measurements $\{G^k_b\}$ and Eve's measurements $\{Z_a\}$  that reproduce the statistics $\vec{p}$. The amount of random bits obtained from Bob's measurements is quantified with the min-entropy of the guessing probability $H_{\min}=-\log_2 G(j,\vec{p})$.

Now let us say more about the additional ninth measurement on Alice's side. We can assume it to be a nine-outcome POVM $\{R_a\}$, which applied on Alice's part of an entangled qutrit state, gives completely random results, i.e. $\Tr[ R_{a} \rho_A ]=1/9$, $\forall a$.
Apart from the above conditions, we require that POVM $\{R_a\}$ should reproduce the statistics given by Eq. \eqref{qcorr}, that is, any Eve's attempt of learning of Alice's outputs remains undetected.
We present an example of a measurement construction meeting the above properties in the section \ref{sec:construction of POVM}.
Our goal is to prove that Eve's guessing probability related to Alice's $j$-th input and consistent with the statistics $\vec{p}$, does not allow Eve to learn anything about Alice's outputs, i.e. $G(j,\vec{p})=1/9$.
Such a situation will provide us $2\log_23$ bits of private randomness. To sum up, we certify randomness based on the correlations, which minimize Eve's guessing probability and these correlations are obtained by measuring POVM $\{R_a\}$ on Alice's subsystem. For that purpose, we employ an arbitrary entangled state of local dimension $3$ certified with the extended Bell scenario and the W-H operators on Bob's side certified with the use of Bell test.

\subsection{Non-local scenario and Bell inequality\label{SecIIB}}

In Ref. \cite{KST+19} authors presented a modification of the Buhrmann-Massar Bell inequality \cite{BM05} to show self-testing of the maximally entangled state of two-qutrits and three mutually unbiased bases on each site. In our work, we apply this certification scenario to self-test the complete set of W-H operators in dimension three. Later this result is used for the certification of randomness from the measurement $F_8=\{R_a\}$. The main idea behind self-testing of the full W-H basis is taken from Refs. \cite{APV+16,WKB+20} and consists in using twice the Bell inequality from Ref. \cite{KST+19}. Below we introduce crucial elements needed to present self-testing results.

Throughout this work, we are using the correlation picture or, equivalently, the observable picture to describe all correlations observed between Alice and Bob, i.e.,  
\begin{equation}\label{ExpValues}
    \langle A_{l|j} B_{m|k} \rangle = \sum^{2}_{a,b=0} \omega^{al+bm} p(ab|jk),
\end{equation}
where $\omega=\exp(2\pi\mathbbm{i}/3)$ and $l,m=0,1,2$. The above formula is a two-dimensional Fourier transform of the conditional probabilities $p(ab|jk)$. Operators $A_{l|j},B_{m|k}$ provide us an alternative description of the measurements $\{F_{a|j}\}$ and $\{G_{b|k}\}$ \cite{KST+19}, and are defined by Fourier transform in the following way
\begin{equation}
    A_{l|j} = \sum^{2}_{a=0} \omega^{al} F_{a|j}, \qquad B_{m|k} = \sum^{2}_{b=0} \omega^{bm} G_{b|k}.
\end{equation}
Let us note here that we make no assumptions about Alice's and Bob's measurements or the shared state; in fact, we consider a fully general situation of $\rho_{AB}$ being mixed and Alice's and Bob's measurement being POVM's. In such a general situation, the above measurement operators $A_{l|j}$ satisfy $A_{l|j}^{\dagger}A_{l|j}\leq \mathbbm{1}$ and 
$A_{l|j}A_{l|j}^{\dagger}\leq \mathbbm{1}$ for any $l$ and $j$, and
$A_{0|j}=\mathbbm{1}$ for any $j$ (the same holds for $B_{m|k}$ operators).

Since in our scenario we are dealing with three-outcome measurements, it is not difficult to observe that $A_{2|j}=A_{1|j}^{\dagger}$. Therefore by also taking into account the fact that $A_{0|j}=\mathbbm{1}$, this implies that measurement is fully determined by a single operator $A_{1|j}$ which, for simplicity, we denote $A_j$; analogously for Bob's measurements, we denote $B_k\equiv B_{1|k}$. In the case of the measurements $F_j$ and $G_k$ being projective, $A_{l|j}$ and $B_{m|k}$ are all unitary operators whose spectra is $1,\omega,\omega^2$, and can be represented as $A_{l|j}=A_j^l$ and $B_{m|k}=B_{k}^m$, where the superscripts $l$ and $m$ are operator powers of unitary quantum observables $A_j$ and $B_k$, such that $A_j^d=B_k^d=\mathbbm{1}$. In this case the expectation values (\ref{ExpValues}) can be expressed as 
\begin{equation}
\langle A_{l|j} B_{m|k} \rangle = \langle\Psi_{ABE}|A_j^l\otimes B_k^m \otimes \mathbb{1} |\Psi_{ABE}\rangle. 
\end{equation}

Let us introduce now the Bell inequality used in our scenario. We consider a slightly simplified Bell operator from \cite{KST+19} for $d=3$, which is sufficient for our purposes. The modification results from the omission of the identity term, and now Bell operator is given by  
\begin{equation}
\label{eq:Wd-definition}
W_1 := \frac{\lambda}{27}\sum^2_{j,k=0} \omega^{jk} A_{j} \otimes B_{k}+h.c.,
%
\end{equation}
where $\lambda=e^{-\mathbbm{i}\pi/18}$ and h.c. stands for the hermitian conjugation.
The corresponding Bell inequality is defined as
\begin{equation}
\label{Bell inequality}
   \langle W_1 \rangle \leq \beta_L, 
\end{equation}
where $\beta_L$ is its classical bound, that is, the maximal value of the Bell expression $\langle W_1 \rangle$ over all correlations admitting the local-realistic description, and it amounts to
\begin{align*}
\beta_{L} &= \frac{ 2 \cos (\pi/9) }{ 3 \sqrt{3} }.
\end{align*}
Moreover, the maximal quantum value of the above Bell inequality 
was found in Ref. \cite{KST+19} to be
\begin{equation}
    \beta_{Q} = \frac{ 2 }{ 3 \sqrt{3} }.
\end{equation}
It is achieved by the maximally entangled state of two qutrits,
\begin{equation}\label{eq:maxent}
\ket{\Phi_{AB}} := \frac{1}{\sqrt{3}} \sum_{j = 0}^{2} \ket{i}_{A} \ket{i}_{B}
\end{equation}
and the following choice of Bob's obesrvables
\begin{equation}
\label{eq:Bk1-definition}
B_0=Z,\quad\, B_1=X,\quad\, B_2=\omega X^2Z^2,
\end{equation}
where
\begin{equation}
X:=\sum_{i=0}^{2}\ket{i+1}\!\!\bra{i} \nbox{and} Z:=\sum_{i=0}^{2} \omega^{i} \ket{i}\!\!\bra{i}
\end{equation}
and $\ket{3}\equiv\ket{0}$. It is worth noticing that the eigenvectors of $B_i$ form mutually unbiased bases in $\mathbbm{C}^3$. At the same time, Alice's optimal observables can be expressed as the following linear combinations of the above optimal observables of Bob,
\begin{equation}\label{def:Alice observables}
A_{j} :=  \frac{ \lambda^{*} }{\sqrt{3}} \sum_{k} \omega^{-jk}B_k^{*},
\end{equation}
where $^{*}$ denotes the complex conjugation in the standard
basis.

As proven in Ref. \cite{KST+19} (see Appendix B therein), the maximal violation of the above Bell inequality is achieved by correlations obtained from the first preparation $\ket{\Psi_{ABE}^1}$. It implies that Bob's measurements $B_i$ with $i=0,1,2$ are projective and that $\dim(\sH_{B})=3 \cdot t_B$ for some positive integer $t_B$, or, equivalently, that $\sH_{B}=(\mathbbm{C}^3)_{B'}\otimes(\mathbbm{C}^{t_B})_{B''}$, and that there exist a unitary operation $U_B:\sH_{B}\rightarrow \sH_{B}$
such that
\begin{align}\label{Bob operators}
\begin{split}
   U_B B_0 U_B^{\dagger} &= Z\otimes Q_1+ Z\otimes Q_2, \\
    U_B B_1 U_B^{\dagger} &= X\otimes Q_1+ X^2\otimes Q_2, \\
    U_B B_2 U_B^{\dagger} &= \omega X^2Z^2\otimes Q_1+  XZ^2\otimes Q_2, 
\end{split}
\end{align}
where operators $Q_{1}, Q_{2}$ are orthogonal projectors satisfying $Q_{1} + Q_{2} = \mathbb{1}_{B''}$. These two projectors identify the orthogonal subspaces corresponding to two inequivalent sets of observables maximally violating the Bell inequality (\ref{Bell inequality}) that are related via transposition. 

Regarding the side of Alice, we can draw a similar conclusion as for Bob's subsystem, as well as we can determine the form of the first preparation $\ket{\Psi_{ABE}^1}$ \cite{KST+19}. However, we will not use their explicit forms in what follows; therefore, we do not present them here.
In fact, the aim of performing 
the Bell test on the first preparation is to certify Bob's measurements.

\subsection{Steering inequality} \label{sec:ext scenario}

In this subsection, we recall the steering inequality introduced in Ref. \cite{SBJ+21}, and we show how to use it for certification of the second preparation. 

Let us again consider Alice and Bob performing measurements on some quantum state $\ket{\Psi_{ABE}^2}$. For a moment, assume that one of the measuring devices, let us say that belonging to Bob, is trusted and performs fixed measurements; Alice's measurement device remains untrusted. Let us then consider a steering inequality constructed in Ref. \cite{SBJ+21},
\begin{equation}\label{Stedupa}
    \langle W_3\rangle\leq \widetilde{\beta}_{L},
\end{equation}
where $W_3$ is a steering operator given by 
\begin{eqnarray}\label{stein}
W_3&=&A_6\otimes B_{0} + \gamma A_7\otimes B_{1}+\delta_1B_0+A_6^{\dagger}\otimes B_{0}^{\dagger}\nonumber\\ 
&&+ \gamma A_7^{\dagger}\otimes B_{1}^{\dagger}+\delta_1^*B_0^{\dagger},
\end{eqnarray}
where
\begin{eqnarray}\label{value1}
\gamma=3\left(\sum_{\substack{i,j=0\\i\ne j}}^{2}\frac{\alpha_i}{\alpha_j} \right)^{-1},\quad \delta_k=-\frac{\gamma}{3}\sum_{\substack{i,j=0\\i\ne j}}^{2}\frac{\alpha_i}{\alpha_j}\omega^{-kj}.
\end{eqnarray}
Coefficients $\gamma$ and $\delta_k$ are functions of three positive numbers $\alpha_i$ such that $\alpha_1^2+\alpha_2^2+\alpha_3^2=1$. Then, $\widetilde{\beta}_L$ is the classical bound of (\ref{Stedupa}).
Whereas we do not know its explicit form for any $\alpha_i$, it was proven in Ref. \cite{SBJ+21} that for any $\alpha_i>0$ it is strictly lower than the maximal quantum 
value of $\langle W_3\rangle$, $\widetilde{\beta}_Q=3$.

Recall also that $\alpha_i$ are Schmidt coefficients of the state 
\begin{eqnarray}\label{partiallyentangledstate}
\ket{\psi(\pmb{\alpha})}=\sum_{i=0}^{2}\alpha_{i}\ket{i_A}\ket{i_B},
\end{eqnarray}
that maximally violates the inequality (\ref{Bell inequality}) 
for observables on the trusted side fixed to be $Z$ and $X$, where we denoted $\pmb{\alpha}=(\alpha_1,\alpha_2,\alpha_3)$. 

Importantly, as shown in Ref. \cite{SBJ+21} from the maximal violation of the steering inequality (\ref{stein}) and the observables $B_0$, $B_1$ certified to be \eqref{Bob operators}, we can device-independently
characterise the second preparation $\ket{\Psi_{ABE}^2}$.
Precisely, we can determine the form of Alice's observables $A_6$ and $A_7$ in the sense that up to some unitary $U_{A}: \sH_{A} \to \sH_{A}$ we have
\begin{equation}
U_A\,A_6\,U_A^{\dagger}=Z\otimes \mathbb{1}_{A''},\quad U_A\,A_7\,U_A^{\dagger}=X\otimes \bar{P}_1+X^2\otimes \bar{P}_2,
\end{equation}
such that $\bar{P}_{1}, \bar{P}_{2}$ are orthogonal projectors satisfying $\bar{P}_{1} + \bar{P}_{2} = \mathbb{1}_{A''}$. Moreover, the joint state corresponding to the second preparation $\ket{\Psi^2_{ABE}}$ up to the unitaries acting on Alice's and Bob's systems can be expressed as
%
\begin{equation}\label{eq:state}
(U_A\otimes U_B\otimes\mathbb{1}_E)\ket{\Psi^2_{ABE}}=\ket{\psi(\pmb{\alpha})}_{A'B'}\otimes\ket{\xi_{A''B''E}}.
\end{equation}
Therefore the expression \eqref{eq:state} constitutes a self-testing statement for any entangled state in dimension three.


\section{Results}

\subsection{Certification of full Weyl-Heisenberg basis in $d=3$ }
\label{sec:H-W basis}

In this subsection we present a method for device-independent 
certification of a full W-H basis, which, up to some phases, consists of the operators $W_{p,q}:=X^pZ^q$, with $p,q=0,1,2$.
We begin by observing that the eigenvectors of the particular subset of the W-H operators $W_{p,q}=X^pZ^q$, that is, $\{Z,X,XZ,XZ^2\}$ are mutually unbiased for $d=3$. Let $M_r$ denote subsequent elements of this subset, were $r=0,1,2,3$, and $\Pi^{(b)}_{r}$ denote the projectors constructed from the eigenvectors of $M_r$, where $b=0,1,2$  labels the outcomes of a given observable. Operators $M_r$ can be written in terms of the spectral decomposition as
\begin{equation}
M_{r}=\sum_{b=0}^2 \omega^b \Pi^{(b)}_{r}. 
\end{equation}
Let us note that multiplying the operators $M_r$ with powers of $\omega$ or taking conjugate transpose results just in relabeling $M_r$'s outcomes.  
Now let us observe that apart from the identity, the remaining W-H operators $\{Z^2, X^2, X^2Z, X^2Z^2\}$ can be obtained from the set $\{M_r\}$ through a suitable rearrangement of the eigenvalues that is
\begin{align}\label{eq:relabaling}
\begin{split}
    Z^2=M^{\dagger}_0&=\sum_b \omega^{-b} \Pi^{(b)}_{0},  \\
    X^2=M^{\dagger}_1&= \sum_b \omega^{-b} \Pi^{(b)}_{1},  \\
    X^2Z=\omega M^{\dagger}_3&= \sum_b \omega^{-b+1} \Pi^{(b)}_{3},  \\
    X^2Z^2=\omega^2 M^{\dagger}_2&= \sum_b \omega^{-b+2} \Pi^{(b)}_{2}. 
\end{split}
\end{align}
The conclusion from the above consideration is that it is enough to certify only $4$ particular observables out of full W-H basis, while the remaining ones can be obtained by adequately relabelling the outcomes. 

As shown in Sec. \ref{SecIIB}, the maximal violation of the Bell inequality (\ref{Bell inequality}) allows one to certify three such particular observables. To certify the fourth one we consider another Bell operator given by
%
%
\begin{align}
\begin{split}
\label{eq:Belloperator2ver2}
W_2 := \frac{\lambda}{27}\Big[ A_3&\otimes \left(B_0+B_2^{\dagger} +B_3 \right) + \\
A_4&\otimes \left(B_0+\omega B_2^{\dagger} + \omega^2 B_3 \right) + \\
A_5&\left.\otimes \left(B_0+\omega^2 B_2^{\dagger} +\omega B_3 \right)\right] +h.c. 
\end{split}
\end{align}
Notice that while Alice's measurements in this Bell operator are all different from 
those used in $W_1$, on Bob's side two measurements $B_0$ and $B_2$ are same.

We aim to prove that observation of the maximum violation of Bell inequality (\ref{Bell inequality}) by the two different sets of observables corresponding to the Bell operators $W_1$ and $W_2$ self-tests the complete W-H basis in dimension three up to the unitary and transposition equivalences. First, 
maximal violation of (\ref{Bell inequality}) by $W_1$ implies that 
Bob's measurements are projective and that 
the corresponding observables $B_i$ with $i=0,1,2$ are of the form (\ref{Bob operators}). Second, it follows from Ref. \cite{KST+19} that maximal violation of the same
inequality by $W_2$ implies that $B_3$ is a quantum observable too and that 
$B_0$, $B_2^{\dagger}$ and $B_3$ must satisfy the following relations:
\begin{align} \label{eq:comrel2}
\begin{split}   
B_0^{\dagger}=-\omega\{B_2^{\dagger},B_3\},\\
B_3^{\dagger}=-\omega\{B_0,B_2^{\dagger}\},\\
B_2=-\omega\{B_3,B_0\}.
\end{split}
\end{align}
These conditions can be used to reconstruct the fourth observable of Bob, $B_3$.
Precisely, plugging the forms of $B_0$ and $B_2$ into the second relation in \eqref{eq:comrel2} one immediately finds that 
\begin{equation}
    U_B B_3 U_B^{\dagger} = \omega^2X^2Z\otimes Q_1+ XZ\otimes Q_2.
\end{equation}

Let us refer here to the fact that we cannot distinguish between optimal measurements used in the device-independent scenario and their transpositions based only on the  correlations $\vec{p}$ observed in a Bell experiment. We call this ambiguity transposition inequivalence. For this reason, we are unable to perform a full tomography of the POVM on Alice's side. In other words, it is not possible to completely reconstruct the elements of the POVM implemented on Alice's side from the joint correlations. Consequently, we cannot certify the maximum randomness directly from the POVM elements. Nevertheless, in the following section, we present a way to overcome this difficulty. We show how to certify maximal randomness from the POVM on Alice's side without certifying the measurement itself.

To summarize this section, we managed to certify four observables on Bob's side by using two Bell operators $W_1$ and $W_2$. Then, we observe that with a proper relabeling \eqref{eq:relabaling} we can construct the other four elements of the W-H basis. With the identity operator, we can then have the complete set of nine W-H operators in dimension three. For the convenience of further calculations, up to certain relabeling of the outcomes, let us express the certified observables $B_k$ as follows 
\begin{eqnarray}\label{CertWeyl}
U_B\,B_{p,q}\,U_B^{\dagger}=X^pZ^q\otimes Q_1+Z^qX^{2p}\otimes Q_2,
\end{eqnarray}
where $p,q=0,1,2$. 


\subsection{Certification of randomness}
\label{sec:randomness}

We can finally proceed to the randomness certification. 
As discussed before, the last ingredient of our scheme is the nine-outcome POVM that Alice  performs on her subsystem. Since the state on which $F_8$ acts is certified to be \eqref{eq:state}, without loss of generality, we can consider this measurement to be a POVM acting on a state of dimension $3\cdot t$ where $t$ is some positive integer. Let us denote the POVM under consideration by $\{\tilde{R}_{a}\}$ where $a=0,1,\dots,8$ represents the outcomes of the measurement. The correlations between the outcomes of nonideal POVM and Bob's observables $B_{p,q}$  (\ref{CertWeyl}) should equal those of the ideal setup, as expressed in Eq. \eqref{qcorr}. This means that 
\begin{align} \label{ExtnonidPOVM}
    \bra{\Psi_{ABE}^2} &\tilde{R}_{a} \otimes B_{p,q} \otimes \mathbb{1}_E \ket{\Psi_{ABE}^2}  \nonumber \\
    &= \bra{\psi(\pmb{\alpha})} R_{a} \otimes W_{p,q} \ket{\psi(\pmb{\alpha})},
\end{align}
where $\{R_a\}$ represents some ideal extremal POVM.  Let us recall that POVMs of a fixed number of outcomes form a convex set, and given POVM  is called extremal if it cannot be decomposed as a convex mixture of other POVMs. We will now demonstrate that the above construction provides the certification of maximal amount of local randomness from the nonideal POVM. 

Let us note here that the ideal POVM elements $R_a$ used in the definition \eqref{ExtnonidPOVM} can be written as 
\begin{equation}\label{idealPOVM}
    R_a=\sum_{k,l=0}^{d-1}r^a_{k,l} P^{-1}\left(X^{k}Z^{l}\right)^*P^{-1},
\end{equation}
which follows from the fact that $R_a$ belongs to a three-dimensional Hilbert space and $P=\sum_{i=0}^{2}\alpha_i\ket{i}\!\!\bra{i}$, such that $\alpha_i\ne 0$ for all $i$. For a remark, let us note that operators $P^{-1}\left(X^{k}Z^{l}\right)^*P^{-1}$ form a complete operator basis for measurements acting on three-dimensional Hilbert space.
Using the fact that $\ket{\psi(\pmb{\alpha})}= \sqrt{3}P\otimes\I_B\ket{\Phi}$, where
\begin{equation}
 \ket{\Phi}=\frac{1}{\sqrt{3}}\left(\ket{00}+\ket{11}+\ket{22}\right),  
\end{equation}
let us now look at the right-hand side of Eq. \eqref{ExtnonidPOVM} which, after a simple computation gives
\begin{equation}\label{eq:rcoef}
    \bra{\psi(\pmb{\alpha})} R_{a} \otimes W_{p,q} \ket{\psi(\pmb{\alpha})}=r^a_{p,q}.
\end{equation}
Hence, from \eqref{ExtnonidPOVM} we have that
\begin{equation}\label{eq:coefititents form POVM}
       \bra{\Psi_{ABE}} \tilde{R}_{a} \otimes B_{p,q} \otimes \mathbb{1}_E\ket{\Psi_{ABE}}=r^a_{p,q}.
\end{equation}
Next we can substitute \eqref{eq:state} into \eqref{eq:coefititents form POVM} and get
\begin{align}
\begin{split}
 \bra{\xi_{A''B''E}}\bra{\psi(\pmb{\alpha})} U_A \tilde{R}_{a} &U_A^{\dagger} \otimes U_B B_{p,q} U_B^{\dagger} \\
  &\otimes \mathbb{1}_E           \ket{\psi(\pmb{\alpha})}\ket{\xi_{A''B''E}},
 \end{split}
\end{align}
where from now on, we will use the notation $\bar{R}_{a}=U_A \tilde{R}_{a} U_A^{\dagger}$, and $U_B B_{p,q} U_B^{\dagger}$ is given by the Eq. \eqref{CertWeyl}.

Since $\{\bar{R}_a\}$ acts on a subsystem of dimension $3\cdot t$, without loss of generality, we can decompose its elements as
\begin{equation}\label{nidPovmDec}
    \bar{R}_{a}=\sum_{k,l=0}^2 P^{-1}\left(X^{k}Z^{l}\right)^*P^{-1} \otimes \bar{R}^a_{k,l},
\end{equation}
where $\bar{R}^a_{k,l}$ act on $\mathcal{H}_{A''}$. Now we can insert expression for $\bar{R}_{a}$ \eqref{nidPovmDec}, the state $\ket{\Psi_{ABE}^2}$ from \eqref{eq:state} and Bob's measurements $B_{p,q}$ \eqref{CertWeyl} into Eq. \eqref{eq:coefititents form POVM} to find the following formulas
\begin{align}\label{eq:coeffitients1}
\bra{\xi_{A''B''E}} \bar{R}^{a}_{0,q} \otimes \mathbb{1}_{B''E} \ket{\xi_{A''B''E}} =r^a_{0,q}
\end{align}
for $p=0$ and all $q$, and 
\begin{align}\label{eq:coeffitients2}
   \bra{\xi_{A''B''E}} &\bar{R}^{a}_{p,q} \otimes Q_1  \otimes \mathbb{1}_E \nonumber \\
    &+\omega^{2pq} \bar{R}^{a}_{2p,q} \otimes Q_2   \otimes \mathbb{1}_E\ket{\xi_{A''B''E}}  =r^a_{p,q}
\end{align}
for $p=1,2$ and all $q$.
Inspired by Ac\'in \textit{et al}. \cite{APV+16}, we define the normalized states 
\begin{equation}
    |\phi^{b,e}_{A''}\rangle= \frac{1}{\sqrt{q_{b,e}}}\left(\mathbb{1} \otimes Q_{b} \otimes \sqrt{Z_e} \right)\ket{\xi_{A''B''E}},
\end{equation}
where $b=1,2$; $Z_e$ is a POVM element corresponding to Eve's outcome $e$ and $\sqrt{q_{b,e}}$ is a normalization factor.
Now by using $|\phi^{b,e}_{A''}\rangle$, we can reformulate coefficients \eqref{eq:coeffitients1} and \eqref{eq:coeffitients2} 
as follows 
\begin{equation}\label{eq:decomposition}
    r^a_{p,q}=\sum_{b=1,2}\sum_{e} q_{b,e} \tilde{r}^{a;b,e}_{p,q}, 
\end{equation}
where we have defined the coefficients 
\begin{equation}\label{Ecoefff1}
\tilde{r}^{a;1,e}_{p,q}:= \bra{\phi^{1,e}} \bar{R}^{a}_{p,q} \otimes \mathbb{1}_{B''} \otimes \mathbb{1}_E  \ket{\phi^{1,e}},
\end{equation}
and
\begin{equation}\label{Ecoefff2}
\tilde{r}^{a;2,e}_{p,q}:=\omega^{2pq}  \bra{\phi^{2,e}} \bar{R}^{a}_{2p,q} \otimes \mathbb{1}_{B''} \otimes \mathbb{1}_E  \ket{\phi^{2,e}}.
\end{equation}
Let us now define the following operators in terms of coefficients \eqref{Ecoefff1}, \eqref{Ecoefff2}
\begin{equation} \label{decEpovm1}
    \bar{R}_{a}^{1,e}=\sum_{p,q}\tilde{r}^{a;1,e}_{p,q} P^{-1}\left(X^{p}Z^{q}\right)^*P^{-1},
\end{equation}
and
\begin{equation} \label{decEpovm2}
    \bar{R}_{a}^{2,e}=\sum_{p,q}  \tilde{r}^{a;2,e}_{p,q} P^{-1}\left(X^{p}Z^{q}\right)^*P^{-1}.
\end{equation}

It is easy to check that the operators $\bar{R}_{a}^{b,e}$  are valid POVM's, that is, they satisfy the following properties, 
$\bar{R}_{a}^{b,e} \ge 0 $ and $\sum_a \bar{R}_{a}^{b,e}  =  \mathbb{1}$. To see this, using Eq. (\ref{Ecoefff1}) let us rewrite the decompositions of the operators given by Eq. (\ref{decEpovm1}) as follows
\begin{equation} 
    \bar{R}_{a}^{1,e}= \Tr_{A''B''E}\left[ (\bar{R}_{a} \otimes \mathbb{1}_{B''E} )  (\mathbb{1}_{A} \otimes \ket{\phi^{1,e}}\!\bra{\phi^{1,e}})\right].
\end{equation}
From the fact that $\bar{R}_{a} \ge 0$ and $\sum_a \bar{R}_{a} =\mathbb{1}$ and from the above equation, it next follows that $\bar{R}^{1,e}_{a} \ge 0$ and $\sum_a \bar{R}^{1,e}_{a} =\mathbb{1}$. Therefore, the coefficients $\tilde{r}^{a;1,e}_{p,q}$ define a family of valid POVMs with the operators $\bar{R}^{1,e}$.
Note that the operators $\bar{R}_{a}^{2,e}$ are transpose of the operators $\bar{R}_{a}^{1,e}$. Thus, the coefficients  $\tilde{r}^{a;2,e}_{p,q}$ also define a family of POVMs with the operators $\bar{R}^{2,e}$.

 Now, using Eq. \eqref{decEpovm1} and Eq. \eqref{decEpovm2}, we have the following expression from Eq. \eqref{eq:decomposition}
\begin{eqnarray}
 R_a=\sum_{b,e} q_{b,e} \bar{R}_{a}^{b,e},
\end{eqnarray}
which can be understood as a convex decomposition of the ideal POVM $\{R_a\}$ in term of the POVMs $\bar R^{b,e}$ with respective weights $q_{b,e}$. But the POVM $\{R_a\}$ is extremal and can not be expressed as convex combination of other POVM's. Thus, we have $\tilde{r}^{a;b,e}_{p,q} =r^a_{p,q}$ for all $b,e$ and $\sum_{b,e} q_{b,e}=1$. 
Finally, rewriting the guessing probability (\ref{LGpr}) of Eve for outcomes of Alice's POVM $\{\bar {R}_a\}$, we have
\begin{align}
G(j=8,\vec{p})&=\sum_a \langle \Psi_{ABE}|\bar R_a \otimes \mathbb{1} \otimes Z_a|\Psi_{ABE}\rangle. 
\end{align}
Using Eqs. (\ref{eq:state}) and (\ref{nidPovmDec}) in the above equation, we arrive at
\begin{align}\label{eq:guessingprob}
G(j=8,\vec{p})&=\sum_a \langle \xi_{A''B''E}|\bar R^a_{0,0} \otimes \mathbb{1} \otimes Z_a|\xi_{A''B''E}\rangle. 
\end{align}
Now, using Eq. (\ref{eq:coeffitients1}), we can simplify \eqref{eq:guessingprob} as follows
\begin{align}\label{gp42}
G(j=8,\vec{p})&= \sum_{b,a} q_{b,a}\, \tilde r_{0,0}^{a;b,a}=\sum_{b,a} q_{b,a}r^a_{0,0},
\end{align}
where we have used the constraint on $\tilde r_{0,0}^{a;b,a}$ as argued in the previous paragraph. Now, choosing an ideal POVM $\{R_a\}$ \eqref{idealPOVM} such that 
\begin{eqnarray}\label{POVMcond1}
    r^a_{0,0}=1/9\quad \forall a, 
\end{eqnarray}
and using it for the guessing probability from Eq. \eqref{gp42}, gives $G(j=8,\vec{p})=1/9$.
This implies that using the scheme mentioned above, Alice can securely generate $-\log_2G=2\log_23$ bits of randomness from any partially entangled two-qutrit state provided that there exist extremal POVM's that satisfy the condition \eqref{POVMcond1}. 
As a final remark here let us note that 
Eq. \eqref{eq:rcoef} for $p,q=0,0$ gives 
\begin{equation}\label{eq:rcoef2}
    \bra{\psi(\pmb{\alpha})} R_{a} \otimes \I \ket{\psi(\pmb{\alpha})}=r^a_{0,0},
\end{equation}
which is equivalent to the expression 
\begin{equation}
    \Tr[R_a\rho_A]=r^a_{0,0},
\end{equation}
where $\rho_A=\Tr_B\left(\ket{\psi(\pmb{\alpha})}\! \bra{\psi(\pmb{\alpha})} \right)$. In the following section we present the construction of the extremal POVM which satisfies the condition \eqref{POVMcond1}.

\subsection{Construction of extremal qutrit POVM} \label{sec:construction of POVM}

D'Ariano and collaborators \cite{APP05} have classified all extremal POVMs with discrete output sets. According to this classification, an extremal POVM with $d^2$ outcomes must necessarily be rank-one, and its elements must be linearly independent.
We also require that POVM elements acting on Alice subsystem give equal probabilities, i.e. $\Tr[R_a\rho_A]=1/9$ for all $a$ and $\rho_A$.
Finding a general class of POVM's that meets all the above conditions for any partially entangled state proved to be a demanding task. Below we present a construction provided in Ref. \cite{SBJ+21} that fulfils the above requirements for a well defined subset of entangled two-qutrit states. 

The POVM elements are given by
\begin{equation} \label{extrPOVM}
 R_{a}:=\lambda_{a} \ket{\alpha_a}\!\!\bra{\alpha_a} ,
\end{equation}
with 
\begin{align} \label{eq:alpha1}
    \ket{\alpha_0}=\ket{0}, \quad  \ket{\alpha_8}=\ket{2},
\end{align}
and for $a=1,\dots,7$,
\begin{eqnarray} \label{eq:alpha2}
    \ket{\alpha_a}=\mu_0\ket{0} + \mu_1&& \exp\left(\frac{2\pi \mathbbm{i} (a-1)  }{7}\right) \ket{1} \nonumber \\
    &&+ \mu_2 \exp\left(\frac{6\pi \mathbbm{i} (a-1)  }{7
    }\right) \ket{2}.
\end{eqnarray}
Coefficients $\lambda_{a}$ are given by
 \begin{align}
     \lambda_0&=\frac{1}{9 \alpha_0^2}, \quad  \lambda_2=\frac{1}{9 \alpha_2^2}, \quad  
      \lambda_a=\frac{1}{7}
      \left(3-\lambda_0-\lambda_2 \right), 
 \end{align}
and $\mu_0,\mu_1,\mu_2$,
\begin{align}
    \mu_0=\sqrt{\frac{1-\lambda_0}{7\lambda_1}}, \quad
    \mu_1=\sqrt{\frac{1}{7\lambda_1}},
    \quad\mu_2=\sqrt{\frac{1-\lambda_2}{7\lambda_1}}.
\end{align}
The above structure imposes that $\alpha_0>1/3$ and $\alpha_2>1/3$. By relabeling the indices in \eqref{eq:alpha1} and \eqref{eq:alpha2} we can generalize the above constraint to the requirement that any two of the $\alpha$'s has to be greater than $1/3$. Using the Monte Carlo method \cite{montecarlo}, we have checked that the above condition satisfies $92,6\%$ of states out of $10^7$ randomly generated. States that are not covered by our construction are weakly entangled. 
Linear independence of the POVM elements \eqref{extrPOVM} can be verified with the condition $\sum_a s_a R_{a}=  \mathbb{0} $, where $\mathbb{0}$ is the zero matrix, satisfied if and only if all $s_a=0$.

\section{Conclusions}
 In our work, we introduced a method for device-independent certification of maximal randomness from pure entangled states in dimension $3$ using non-projective measurements. For this purpose, we exploited the extended Bell scenario introduced recently in Ref. \cite{SBJ+21} which combines DI certification of local measurements with one-sided device-independent certification of pure entangled states. In fact, we first showed how to certify the complete set of W-H operators in dimension three by using the self-testing scheme proposed in Ref. \cite{KST+19}. Then, by using the steering scenario proposed in \cite{SBJ+21}, we provide certification of any entangled state in dimension three. These two components finally allowed us to achieve the main goal of the paper, that is, to certify $2\log_23$ random bits by performing generalized measurements on Alice's subsystem of partially-entangled two qutrit states. In fact, we verified numerically that the constructed POVM that we use in our scheme covers a significant subset of $92,6\%$ of bipartite entangled states.


Several interesting directions for further research emerge from our work. 
First, it would be extremely interesting to understand what is the maximal 
amount of global randomness that can be certified from entangled quantum states of
local dimension $d$ by performing non-projective measurements on both subsystems. Whereas the known theoretical limit says that at most $4\log_2d$ bits of randomness can be created in this way, it is unclear whether this limit is achievable. In fact, 
it was shown recently in Ref. \cite{WKB+20} that in the case of two-qubit systems $3.9527$ bits of randomness is actually the maximal amount that can be generated in a device-independent way. This intriguing observation comes from the fact that an adversary can use some of the global correlations obtained with non-projective qubit measurements to infer information about the outcomes of a Bell experiment. Therefore understanding the fundamental limit for generating global randomness from two-qudit states in the non-local scenario remains a very interesting challenge. Another interesting problem for further research is to provide constructions of $d^2$-outcome extremal POVMs that can be used to generate randomness from any, even arbitrarily little entangled states. 





\section*{Acknowledgement}

 
This work is supported by Foundation for Polish Science through the First Team project (no First TEAM/2017-4/31). J.J.B. acknowledges financial support from the National Science Center within the grant Sonatina 4, No. 2020/36/C/ST2/00592.


%

\end{document}